# 1-GHz mid-infrared frequency comb spanning 3 to 13 µm

Nazanin Hoghooghi[1,*], Sida Xing[2,3], Peter Chang[2,3], Daniel Lesko[2,4], Alexander Lind[2,3], Greg Rieker[1], and Scott Diddams[2,3,5,*]

[1]Precision Laser Diagnostics Laboratory, University of Colorado, Boulder, Colorado 80309 USA
[2]Time and Frequency Division, National Institute of Standards and Technology, Boulder, Colorado 80305, USA
[3]Department of Physics, University of Colorado, Boulder, Colorado 80309, USA
[4]Department of Chemistry, University of Colorado, Boulder, Colorado 80309, USA
[5]Electrical Computer and Energy Engineering, University of Colorado Boulder, Colorado 80309, USA
Author e-mail address: *nazanin.hoghooghi@colorado.edu, *scott.diddams@nist.gov

## Abstract

Mid-infrared (MIR) spectrometers are invaluable tools for molecular fingerprinting and hyper-spectral imaging. Among the available spectroscopic approaches, GHz MIR dual-comb absorption spectrometers have the potential to simultaneously combine the high-speed, high spectral resolution, and broad optical bandwidth needed to accurately study complex, transient events in chemistry, combustion, and microscopy. However, such a spectrometer has not yet been demonstrated due to the lack of GHz MIR frequency combs with broad and full spectral coverage. Here, we introduce the first broadband MIR frequency comb laser platform at 1 GHz repetition rate that achieves spectral coverage from 3 to 13 µm. This frequency comb is based on a commercially available 1.56 µm mode-locked laser, robust all-fiber Er amplifiers and intra-pulse difference frequency generation (IP-DFG) of few-cycle pulses in $\chi^{(2)}$ nonlinear crystals. When used in a dual comb spectroscopy (DCS) configuration, this source will simultaneously enable measurements with µs time resolution, 1 GHz (0.03 cm$^{-1}$) spectral point spacing and a full bandwidth of >5 THz (>166 cm$^{-1}$) anywhere within the MIR atmospheric windows. This represents a unique spectroscopic resource for characterizing fast and non-repetitive events that are currently inaccessible with other sources.

## Introduction

Coherent MIR (3-25 µm) light sources are critical to the advancement of various scientific fields. This is particularly true for spectroscopic sensing and imaging, where such sources access the molecular "fingerprint" region (6.7–20 µm), enabling chemical specificity while also improving the minimum detection sensitivity limit. Spectroscopy and imaging systems using CW MIR lasers have shown unprecedented sensitivity[1,2]. Broadband MIR optical frequency combs[3] can further enhance the performance of spectroscopic and imaging systems by offering three important characteristics: high brightness, full instantaneous spectral coverage, and high spectral resolution. When combined with a fast, broadband, and high resolution detection scheme such as dual-comb spectroscopy (DCS), MIR frequency comb spectrometers have the potential for recovering full spectral fingerprint information at MHz rates. A significant body of MIR frequency comb spectroscopy employs dispersive and Fourier transform spectrometers [4–6], but here we restrict our attention to DCS with its simplicity that stems from a single-element detector and freedom from mechanical delay stages.

The majority of existing broadband MIR frequency combs are generated through nonlinear down-conversion of near-infrared (NIR) frequency combs, either through parametric oscillation or difference frequency generation (DFG) techniques. These sources typically have 50-200 MHz comb tooth spacing, which is defined by the repetition rate of the fundamental NIR frequency combs driving the nonlinear process. Such ~100 MHz repetition rate MIR frequency combs have been used in a DCS configuration for atmospheric sensing[7], trace gas detection[7-11], and studies related to wildfire spread[12], where the time scale of the events under study is on the order of seconds. However, a wide range of scientific studies, such as those in combustion[13-15] and biological reactions[16], would benefit from MIR DCS systems with increased speed (µs time resolution), while maintaining broad spectral coverage and high spectral resolution.

Since the measurement speed of DCS scales as the square of the repetition rate[17], significant gains are achieved by scaling broadband MIR frequency combs from the MHz range to the GHz. In particular, frequency combs with ~1 GHz repetition rate strike an attractive balance between speed and spectral resolution in scenarios of expanding interest. For example, they enable DCS measurements over ~5 THz of spectral bandwidth with 10 µs time resolution, while still providing the necessary spectral resolution for gas phase measurements from atmospheric to combustion and exoplanet relevant temperatures and pressures. These benefits were highlighted previously in the near infrared[18], but have not been fully extended to the MIR, where only a handful of 1 GHz MIR frequency combs exist[19-24]. It should be noted that the multi-GHz chip-based MIR frequency combs generated from electro-optic combs[25] quantum cascaded lasers[26] or microcombs[27] enable sub-microsecond spectral acquisition. However, most of these sources have relatively narrow spectral coverage and comb tooth spacing of >10 GHz, which limits their applications in gas phase spectroscopy without tuning the comb offset or repetition rate.

Addressing these challenges, we demonstrate the first 1-GHz MIR frequency comb with spectral coverage from 3 to 13 µm. A key aspect of this advance is the use of soliton self-compression in highly nonlinear fiber (HNLF) to generate NIR pulses centered at 1.56 µm with average power of 2.3 W and duration as short as 8.1 fs (1.5 optical cycles). In a simple single-pass geometry, these ultrashort pulses drive intra-pulse difference frequency generation (IP-DFG) in $\chi^{(2)}$ nonlinear crystals, yielding MIR powers as high as 6.2 mW. Our approach is built off a commercial 1.56 µm source[28] and established Er-fiber amplifiers and fiber components, all of which combine to provide a robust, reproducible, and broad bandwidth MIR frequency comb platform for high-speed molecular spectroscopy in settings beyond the research lab.

## Results

### 1 GHz Intra-Pulse Difference Frequency Generation

There are four main steps in our approach to generate broad bandwidth MIR combs at 1 GHz: (1) Start with a robust 1 GHz mode-locked laser at 1.56 µm; (2) Amplification in Er-doped fiber; (3) Temporal compression and spectral broadening in HNLF; and (4) IP-DFG in a $\chi^{(2)}$ nonlinear crystal. Figure 1 shows the experimental setup (top), together with the evolution of the optical spectrum after each step in the process (bottom). The mode-locked laser is a commercially available source. It generates ~300 fs sech$^2$ pulses with 60 mW of power and 13 nm of bandwidth. We have previously shown the full stabilization of this laser for frequency comb operation[28]. The pulses from the mode-locked laser are first pre-amplified in a core-pumped Er-doped fiber amplifier whose output provides the seed for a high-power chirped-pulse amplifier. After the first stage of amplification, the pulses are temporally stretched in dispersion compensating fiber (DCF) to a few picoseconds. The stretched pulses then seed a cladding-pumped erbium/ytterbium co-doped fiber amplifier (EYDFA)[29]. We designed the pre-amplifier to achieve a factor of 3.5 increase in spectral bandwidth through self-phase modulation (SPM), which in turn compensates the gain narrowing in the EYDFA. The pre-amplifier is seeded with 24 mW of power and the average power after the chirped pulse EYDFA amplifier is 4.3 W corresponding to 4.3 nJ pulse energies (23 dB of net gain). The laser and all fiber components are polarization maintaining (PM).

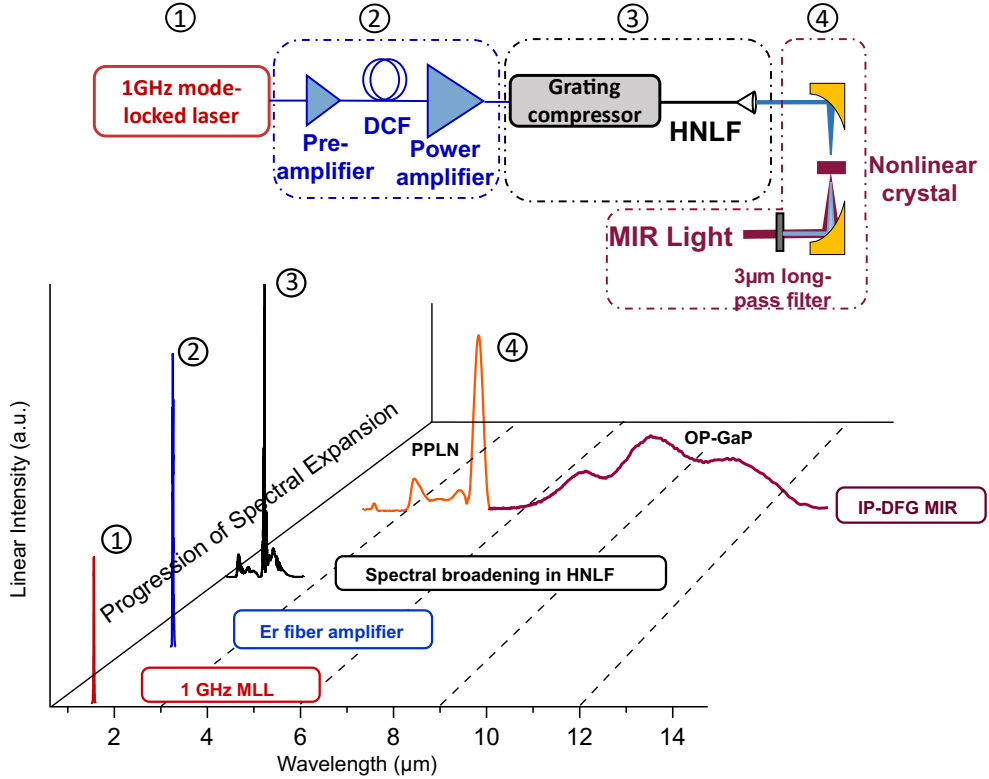

Figure 1. (Top) Schematic of the experimental setup showing the 4 main steps of the 1 GHz IP-DFG MIR frequency comb. The first step is a 1 GHz mode-locked laser centered at 1.56 µm. The second is all-fiber amplification first in a core-pumped EDFA and then in a cladding-pumped EYDFA. The third step is pulse compression in a grating compressor and subsequent spectral broadening in HNLF. And the fourth step is MIR generation through IP-DFG in a $\chi^{(2)}$ non-linear crystal. (Bottom) The evolution of the optical spectrum in each step.

After the amplification, the 4.3 nJ pulses are temporally compressed using a grating compressor to ~120 fs with total efficiency of 84%. The compressed pulses are then spectrally broadened in HNLF[30]. We explored the generation of short 1 GHz pulses using HNLF with both normal and anomalous dispersion. Few-cycle pulses generated in these two regimes are characterized using an all-reflective second-harmonic frequency-resolved optical gating (SHG-FROG). The details of the few-cycle pulse generation and characterization are given in the next two sections. The last step in the system is the MIR generation through IP-DFG where the short pulses are achromatically focused into a $\chi^{(2)}$ nonlinear crystal using an off-axis parabolic mirror with 2-inch focal length. We use the IP-DFG technique instead of the traditional "two-branch" DFG[31,32] since it generates a phase stable output in a compact and single pass setup without the need for delay stage stabilization. In the IP-DFG approach, DFG happens between the frequency components of the same pulse. While in a two-branch DFG, the light from the NIR source is split into two branches, the signal and pump pulses travel different paths and need to be precisely overlapped in the nonlinear crystal (both temporally and spatially) which adds relative intensity noise and complexity[33]. The generated MIR light is collected and collimated using another 2-inch off-axis parabolic mirror. A long-pass Ge filter (>3 µm) is used to separate the generated MIR light from the fundamental and visible light generated through cascaded $\chi^{(2)}$ processes. We studied MIR generation with 3 different nonlinear crystals, periodically poled lithium niobate (PPLN), Cadmium Silicon Phosphide (CSP) and orientation patterned gallium phosphide (OP-GaP). The resulting spectra and optical power generated in each case are given in following sections.

## MIR generation from 7 to 13 µm

For the 7-13 µm range, we use normal dispersion HNLF (ND-HNLF) for spectral broadening in Step 3. Broadening in ND-HNLF enables the generation of a broadband spectrum with a high degree of coherence, a low timing jitter between different wavelengths, and a flat spectrum that can be easily compressed to a clean short pulse using bulk fused silica. Generation of few-cycle NIR pulses at 100 MHz repetition rate using ND-HNLF and subsequent broadband MIR to LWIR frequency comb generation using IP-DFG was successfully implemented in[34,35]. Here, we use a non-polarization maintaining (non-PM) ND-HNLF with D=-1.0 ps/nm-km, for spectral broadening. Pulses from the grating compressor are coupled into a 21-cm-long ND-HNLF. The output spectrum covers 1.3 µm to 1.7 µm. Increasing the length of the ND-HNLF did not yield a broader spectrum due to dispersive broadening of the pulse. After recompression in bulk fused silica glass, which compensates for the positive chirp accumulated in ND-HNLF, we achieve pulses of 21.5 fs duration. The pulses are characterized using SHG-FROG with reconstruction error less than 1%. The retrieved temporal profile of the pulse and the experimental and reconstructed SHG-FROG data are shown in Fig. 2 (a)-(c). The NIR spectrum of the pulse and spectral phase are shown in the inset of Fig. 2 (a).

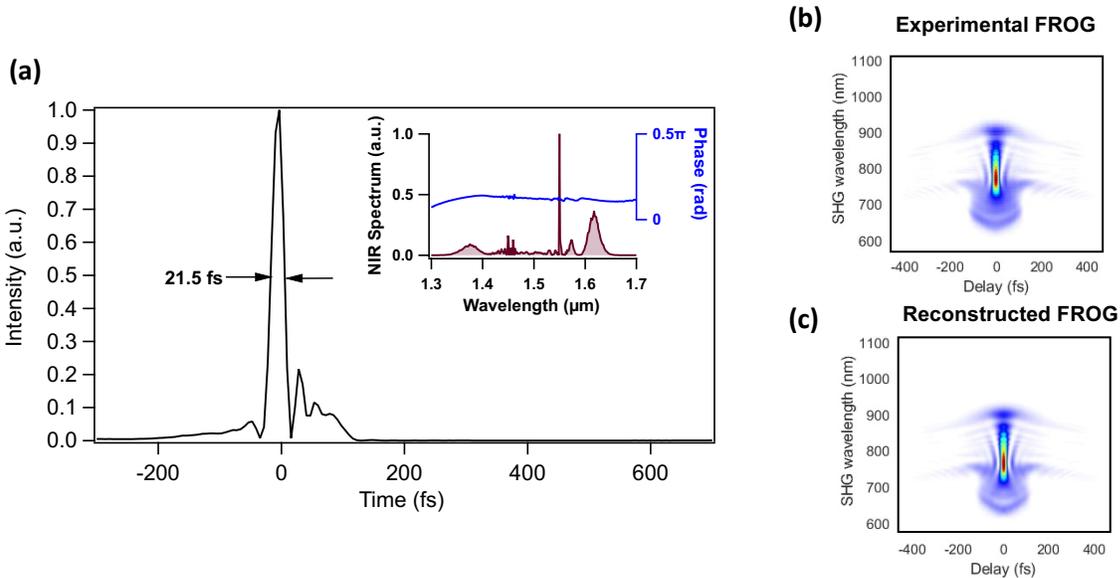

Figure 2. (a) The intensity profile of the short NIR pulse generated with ND-HNLF. SHG- FROG is used to measure the pulse width of 21.5 fs. (Inset) The NIR spectrum of the measured pulse covering 1.3 µm to 1.7 µm and relatively flat spectral phase. (b) Experimental FROG and (c) Reconstructed FROG with < 1% error.

By focusing the 21.5 fs pulse onto a 560-µm-thick CSP crystal we generate MIR spectrum from 7.5 µm to 13 µm. By simply changing the crystal to an OP-GaP crystal (1-mm-thick, orientation patterning period of 61.1 µm) a flatter and broader MIR spectrum from 7 µm to 14 µm is obtained. The optical power generated for both cases is ~100 µW. The MIR spectrum from CSP and the octave spanning spectrum from OP-GaP are shown in Fig. 3. The spectra are measured using a Fourier transform spectrometer (FTS) with 4 cm$^{-1}$ resolution.

Since the short NIR pulse of Fig. 2 has a 45 THz bandwidth, it is not possible to use it to generate MIR wavelengths shorter than 6.5 µm (specifically 3 - 5 µm) through IP-DFG. Extending the spectrum to 1 µm in ND-HNLF requires still higher pulse energies, which are difficult to achieve due to the high repetition rate of the source. The next section shows the use of nonlinear fiber with anomalous dispersion for spectral broadening and subsequent MIR light generation in the 3 - 5 µm range.

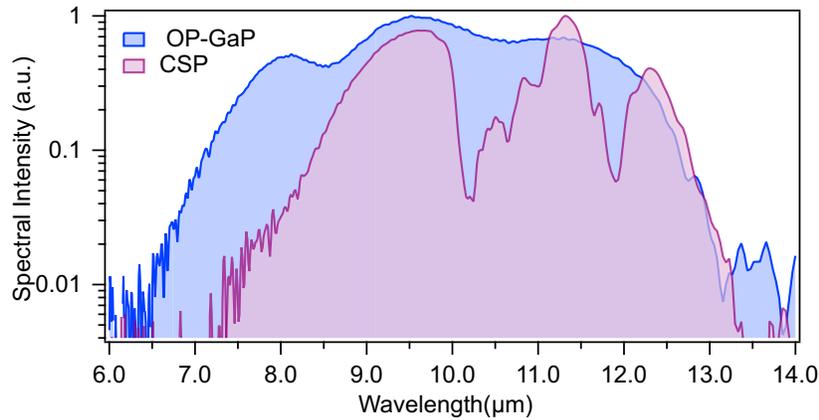

Figure 3. The MIR spectra generated through IP-DFG in a CSP crystal and an OP-GaP crystal.

## MIR generation from 3 to 5 µm

Using soliton self-compression in an anomalous dispersion HNLF (AD-HNLF) we generate sub-two-cycle NIR pulses at 1 GHz directly out of fiber and without any need for further recompression[36]. The pulses after the grating compressor are coupled into a short piece of PM1550 fiber which is spliced to a polarization maintaining AD-HNLF. Here, we use a polarization maintaining AD-HNLF with a dispersion value of D= 5.4 ps/nm-km at 1.55 µm. By cutting back the fiber, we determined the exact length (3.6 cm in this case) of AD-HNLF that generates a broad spectrum covering 1 µm to 2.2 µm. Figure 4 shows the self-compressed pulses measured at the output of AD-HNLF with SHG-FROG (retrieval error of <2%). The measured pulse width is 8.1 fs (Fig.4 (a)) or 1.5 optical cycles, which is, to the best of our knowledge, the shortest reported 1.56 µm pulses directly from an all Er-fiber amplification system. Previous work on few cycle pulse generation directly from an Er-amplification system has shown 9.4 fs pulses at 100 MHz repetition rate[37]. Other notable work is a single cycle Er-fiber system which unlike our single-branch approach, uses a two-branch design of splitting the pulse spectrum, recompressing each pulse and recombining to achieve a single cycle pulse of 4.3 fs[38].

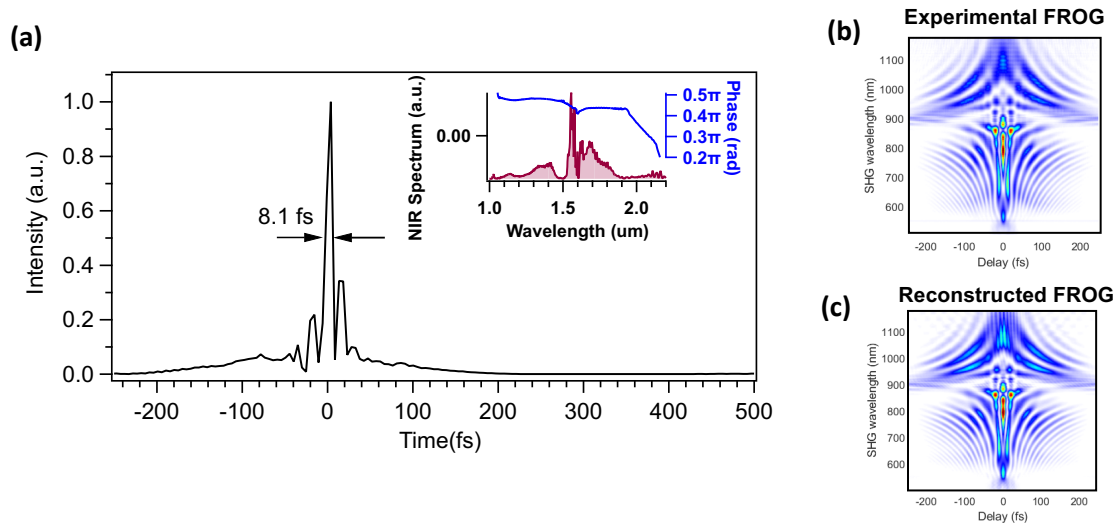

Figure 4. The intensity profile of the sub-two-cycle NIR pulse generated through soliton self-compression in AD-HNLF and measured using SHG- FROG. The measured pulse width is 8.1 fs. (Inset) The NIR spectrum of the measured pulse spanning 1 μm to 2.2 μm. (b) Experimental FROG and (c) Reconstructed FROG with < 2% error.

The 8.1-fs pulse is focused onto a 1 mm long fanout PPLN crystal with grating periods ranging from 27.5 μm to 31.6 μm. The resulting offset-free MIR frequency comb covers the 3 - 5 μm region without gaps and with a maximum power of 4.5 mW. Figure 5 (a) shows the MIR spectra generated by translating the fanout PPLN crystal across the laser beam, thereby changing the poling period. The corresponding optical power for each spectrum is listed above each curve. The dip at 4.2 μm is due to atmospheric $CO_2$ absorption in the free space portions of the comb and Fourier transform spectrometer. By replacing the crystal with another 1 mm long PPLN with 16 periodically poled gratings, we achieve even higher optical power of 6.2 mW covering 3 - 4 μm (the 3 μm cut-off is the Ge long pass filter cut-off). The spectra generated from this crystal when scanning through a selection of the different poling periods and the corresponding MIR power are shown in Fig. 5(b). The combination of broad and full 3-5 μm bandwidth, optical power levels well above the saturation of MIR detectors, ability to tailor the spectrum, and 1 GHz repetition rate make this a very unique source, suitable for a wide range of high-speed DCS applications. Other MIR frequency combs covering the entire 3 - 5 μm region have ∼ 10 times lower repetition rates[37] with some being OPO systems[39].

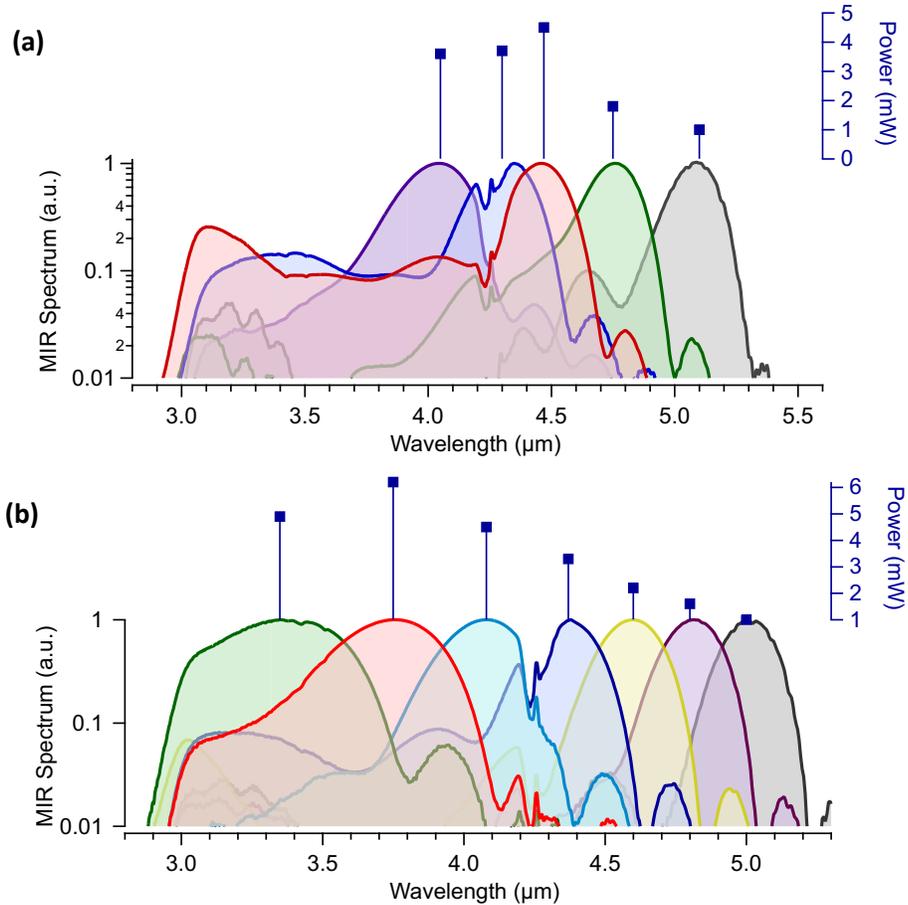

Figure 5. (a) MIR spectra generated by focusing 1.5 cycle pulse into a fanout PPLN crystal. Broadest spectrum covers 3 -4.7 μm and 4.5 mW of power. (b) MIR spectra with a PPLN with 7 individual periodically poled gratings (out of 16 total). Maximum power is 6.2 mW and covers 3 - 4 μm.

## Full frequency stabilization of the comb

An advantageous aspect of driving the IP-DFG process with ultrashort pulses is that cascaded $\chi^{(2)}$ nonlinear processes inside the PPLN crystal give rise to the generation of an $f_{ceo}$ beat note in different wavelength regions of the spectrum[35]. For example, we detect an $f_{ceo}$ beat note at ~ 3.5 μm which is generated through cascaded processes of the DFG between ~1.08 μm (doubled 2.16 μm) and fundamental 1.56 μm light. We were also able to detect the $f_{ceo}$ beat note at wavelengths of ~600 nm and ~900 nm. Figure 6 (a) shows the $f_{ceo}$ beat note detected in the MIR with > 30 dB SNR (RBW 100 kHz), and we use this MIR beat to stabilize the offset frequency of the comb. In addition, we lock an optical comb line of the NIR comb to a narrow linewidth reference laser at 1550 nm. With both the $f_{ceo}$ and the 1550 nm beats phase locked, the entire comb (NIR and MIR) is stabilized. The phase locking is accomplished using a cost-effective, computer-controlled servo system based on a field-programmable gate array (FPGA)[40,41]. Such FPGA-based servo systems, with an open-source and user-friendly computer interface, significantly simplify phase locking of frequency combs and provide a robust and portable platform that facilitates portability of the comb.

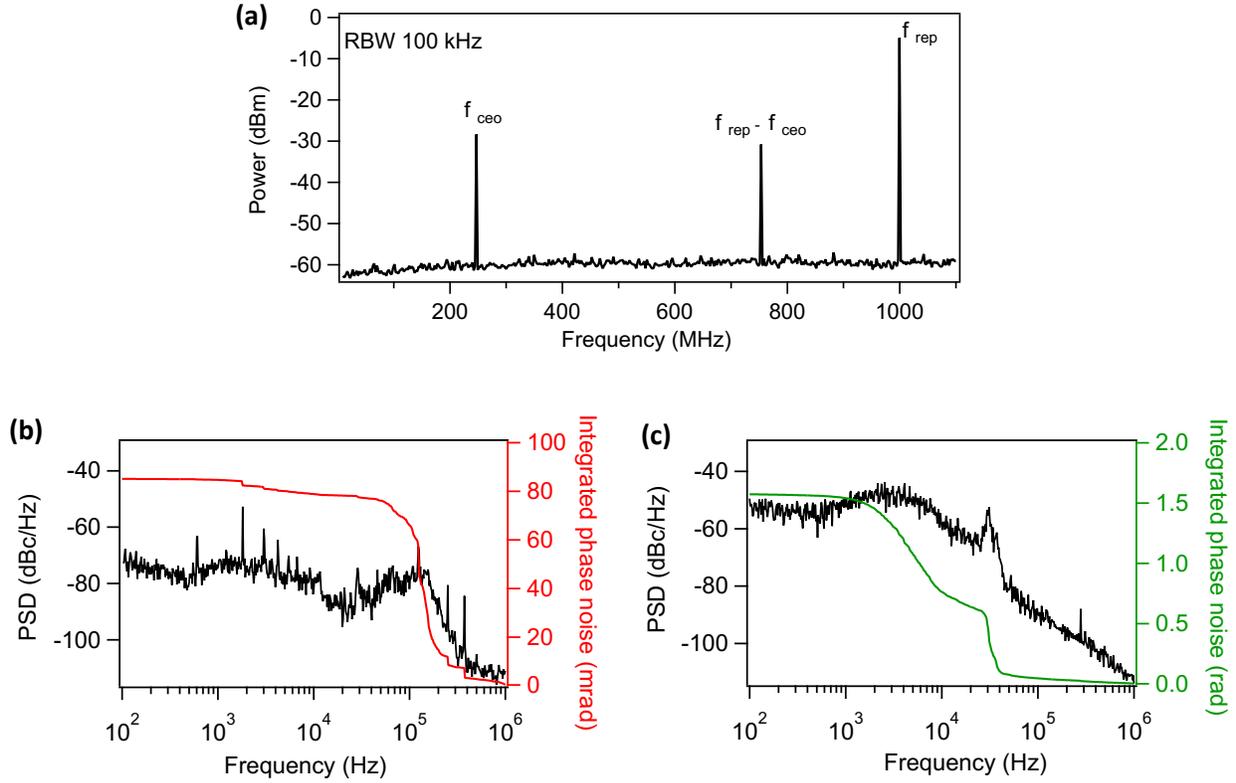

Figure 6. (a) The $f_{ceo}$ beat note generated as a result of cascaded $\chi^{(2)}$ processes in PPLN and detected in the 3.5 μm region. The beat note has an SNR of ~ 30 dB in 100 kHz resolution bandwidth (RBW). (b) Power spectral density plot of the locked optical beat note of a comb tooth at 1.56 μm with a stabilized CW laser showing integrated phase noise of 85 mrad (integrated from 100 Hz to 1 MHz) (c) Power spectral density of the locked $f_{ceo}$ beat note detected in MIR. The integrated phase noise is 1.57 rad (from 100 Hz to 1 MHz)

The measured power spectral density and the integrated phase noise for the $f_{ceo}$ and $f_{beat}$ locks are shown in Fig.f 6 (b) and (c). When integrated from 100 Hz to 1 MHz, the phase noise of the locked optical beat ($f_{beat}$) and the locked offset frequency ($f_{ceo}$) are 85 mrad and 1.57 rad, respectively. This level of integrated noise of $f_{ceo}$ is 3 times higher than we achieved using analog electronics and an $f_{ceo}$ beat detected in the NIR using the *f-2f* technique[28]. However, this value is still well within the range of stability required for spectroscopy applications envisioned for this system. The locks are routinely maintained over an entire day of operation. Using the $f_{ceo}$ detected in the MIR or visible wavelengths for stabilization removes the need for an additional EDFA and *f-2f* setup for detecting $f_{ceo}$ in the NIR. This in turn results in a simpler and more cost-effective system.

## Discussion

We developed the first 1-GHz MIR frequency combs based on a 1-GHz 1.56 μm mode-locked laser, all-fiber amplifiers and robust IP-DFG technique. The MIR spectrum provides full spectral coverage in the important atmospheric windows and molecular fingerprint region from 3 to 13 μm. Unique to our system is the full coverage at a 1 GHz repetition rate from 3 - 5 μm using PPLN and 6.5 - 14 μm using OP-GaP with total power as high as 6.2 mW. Key to generating the broad spectra in the 3 - 5 μm region is the use of NIR driving pulses as short as 8.1 fs (1.5 optical cycles) that arise from soliton self-compression in AD-HNLF. We also find that spectral broadening in ND-HNLF is more appropriate to extend the wavelength coverage to longer wavelengths. We have fully stabilized the frequency comb by phase locking both the repetition rate, via stabilization against a CW laser, and the carrier envelop offset frequency, which conveniently originates from the same nonlinear crystal producing the IP-DFG. We used a simple and cost-effective FPGA-based servo system for phase locking.

Overall, this system has the following important and notable characteristics: 1) broad bandwidth in MIR with continuous coverage from 3 to 5 µm and 7 to 13 µm; 2) 1-GHz repetition rate, enabling fast dual-comb measurements with high spectral resolution in the future; 3) a robust design using commercially available mode-locked lasers, all-fiber Er amplifiers, and the IP-DFG technique (a single-pass and single-pulse approach with inherent robustness and low noise). Together, these advances will enable access to a previously inaccessible regime of high-speed and broad bandwidth dual-comb spectroscopy and hyper-spectral imaging of fast non-repetitive events.

## Acknowledgements

Thanks to NSF 2019195, Defense Advanced Research Project Agency (W31P4Q-15-1-0011), Air Force Office of Scientific Research (FA9550-20-1-0328) and NIST for funding as well as Daniel Herman, Kristina Chang, Franklyn Quinlan and Ian Coddington for their contributions and comments on this manuscript.